\documentstyle[12pt,aps,prd,preprint,graphicx]{revtex}
\begin{document}
\title{Late-Time Tails in Gravitational Collapse of a
Self-Interacting (Massive) Scalar-Field and Decay of a 
Self-Interacting Scalar Hair}
\author{Shahar Hod and Tsvi Piran}
\address{The Racah Institute for Physics, The
Hebrew University, Jerusalem 91904, Israel}
\date{\today}
\maketitle

\begin{abstract}
We study {\it analytically} the initial value problem for a {\it
self-interacting} (massive) scalar-field on a Reissner-Nordstr\"om
spacetime.  Following the no-hair theorem we 
examine 
the {\it
dynamical} physical mechanism by which the {\it self-interacting }
(SI) hair decays. We show that the intermediate asymptotic behaviour
of SI perturbations is dominated by an oscillatory inverse
power-law decaying tail.
We show that at late-times the decay of 
a SI hair is {\it slower} than any power-law.
We confirm our {\it analytical} results by numerical simulations.
\end{abstract}

\section{introduction}\label{Sec1}
The late-time evolution of various fields outside a collapsing star has an
important implications 
to 
two major aspects of black-hole physics:
the no-hair theorem and the mass-Inflation scenario. 
The {\it no-hair theorem}, introduced by Wheeler in the early 1970s,
states that the {\it external} field of a black-hole relaxes to a
Kerr-Newman field characterized solely by the black-hole's mass,
charge and angular-momentum. Thus, it is of interest to 
explore 
the {\it dynamical} physical mechanism responsible for the relaxation of
perturbations fields outside a black-hole and to determine the
decay-{\it rates} of the various perturbations (which {\it differ}
from one field to the other).  The mechanism by which massless neutral
fields are radiated away was first studied by Price
\cite{Price}. 
The physical mechanism by which a massless charged scalar hair is radiated
away was studied in \cite{HodPir1,HodPir2}. In this paper we study the
physical mechanism responsible for the decay of a 
{\it self-interacting (massive)} scalar-hair.

The asymptotic late-time tails along the outer horizon of a rotating
or a charged black-hole are used as initial input for the ingoing
perturbations which penetrates into the black-hole. These
perturbations are the physical cause for the well-known phenomena of
{\it mass-inflation}
\cite{Poisson}. 
In this context,
one should take into account the existence of {\it massive}
tails outside the collapsing star. 
Here we study analytically the intermediate asymptotic behaviour of such self-interacting
(massive)  perturbations fields. We study the late-time 
asymptotic behaviour numerically and confirm the numerical results of
Burko \cite{Burko}.

The plan of the paper is as follows. In Sec. \ref{Sec2} we 
describe 
the physical system and formulate the evolution equation
considered. In Sec. \ref{Sec3} we formulate the problem in terms of
the black-hole Green's function using the technique of spectral
decomposition (this section is analogous to the one given in
\cite{HodPir2}).  In Sec. \ref{Sec4} we study the intermediate
asymptotic evolution
of SI scalar perturbations on a Reissner-Nordstr\"om background. We
find an {\it oscillatory} inverse {\it power-law} behaviour of the
perturbations at a fixed radius and along the black-hole
outer-horizon. We find that the dumping exponents which describe the
intermediate fall-off of {\it SI} perturbations are {\it smaller} compared with the
massless (neutral) dumping exponents.  In Sec. \ref{Sec5} we verify
our {\it analytical} results by numerical simulations.  We conclude in
Sec. \ref{Sec6} with a brief summary of our results and their
implications.

\section{Description of the system}\label{Sec2}

We consider the evolution of {\it SI} scalar 
perturbation 
fields
outside a collapsing star.  The external gravitational field of a
spherically symmetric collapsing star of mass $M$ and charge $Q$ is
given by the Reissner-Nordstr\"om metric
\begin{equation}\label{Eq1}
ds^2=-\left( {1-{{2M} \over r}+{{Q^2} \over {r^2}}} \right)dt^2+\left( {1-{{2M}
\over r}+{{Q^2} \over {r^2}}} \right)^{-1}dr^2+r^2d\Omega ^2\  .
\end{equation}
Using the tortoise radial coordinate $y$, defined by $dy=dr/\lambda^2$
where $\lambda^2={1-{{2M} \over r}+{{Q^2} \over {r^2}}}$, the metric becomes
\begin{equation}\label{Eq2}
ds^2=\lambda^2 (-dt^2+dy^2)+r^2d\Omega ^2\  ,
\end{equation}

The wave equation for the SI scalar-field is
\begin{equation}\label{Eq3}
\phi_{;ab}g^{ab} -U'(|\phi|^{2}) \phi =0\  ,
\end{equation}
where $U(|\phi| ^{2})$ is the self-interaction potential and
$U'(|\phi| ^{2})=dU(|\phi|^{2})/d \phi^{*}$.  Since we study the
evolution of 
small 
perturbations 
we 
approximate $U'(|\phi|^{2})$
by $m^{2} \phi$ (we assume that $m$ is real), neglecting terms of
higher order 
in $\phi$.  
Resolving the field into spherical
harmonics $\phi =\sum\limits_{l,m} {\psi _m^l\left( {t,r}
\right)Y_l^m{{\left( { \theta ,\varphi} \right)} \mathord{\left/ 
{\vphantom {{\left( {  \theta ,\varphi } \right)} r}}\right.}r}}$
one obtains a wave-equation for each multiple moment
\begin{equation}\label{Eq4}
\psi _{,tt}-\psi _{,yy}+V\psi =0\  ,
\end{equation}
where

\begin{equation}\label{Eq5}
V=V_{M,Q,l,m}\left( r \right)=\left( {1-{{2M} \over r}+{{Q^2} \over {r^2}}}
\right)\left[ {{{l\left( {l+1} \right)} \over {r^2}}+{{2M} \over {r^3}}-{{2Q^2}
\over {r^4}}+m^{2}} \right]\  .
\end{equation}

\section{Formalism}\label{Sec3}

The time-evolution of a SI scalar-field described by Eq. (\ref{Eq4})
is given by
\begin{equation}\label{Eq6}
\psi (y,t)= \int \left[ G(y,x;t) \psi _t(x,0)+G_t(y,x;t) \psi (x,0)\right] 
dx\ ,
\end{equation}
for $t>0$, where the (retarded) Green's function $G(y,x;t)$ is defined
as
\begin{equation}\label{Eq7}
\left [{{\partial ^2} \over {\partial t^2}} -{{\partial ^2} \over 
{\partial y^2}} +V(r) \right ] G(y,x;t)=
\delta (t) \delta(y-x)\  . 
\end{equation}
The causality condition gives us the initial condition $G(y,x;t)=0$
for $t \leq 0$.  In order to find $G(y,x;t)$ we use the Fourier
transform
\begin{equation}\label{Eq8}
\tilde G(y,x;w)= \int_{0^-}^{\infty} G(y,x;t) e^{iwt} dt\  .
\end{equation}
The Fourier transform is analytic in the upper half $w$-plane and 
it 
satisfies
\begin{equation}\label{Eq9}
\left ({{d^2} \over {dy^2}} +w^{2} -V \right) \tilde G(y,x;w)
= \delta(y-x)\  .
\end{equation}
$G(y,x;t)$ itself is given by the inversion formula
\begin{equation}\label{Eq10}
G(y,x;t)={1 \over {2 \pi}} \int_{- \infty +ic}^{\infty +ic} \tilde G(y,x;w)
e^{-iwt} dw\  ,
\end{equation}
where $c$ is some positive constant.

Next, we define two auxiliary functions $\tilde \psi_1(y,w)$ and 
$\tilde \psi_2(y,w)$ which are 
linearly independent 
solutions to the 
homogeneous equation
\begin{equation}\label{Eq11}
\left ({{d^2} \over {dy^2}} +w^{2} -V \right) \tilde 
\psi_i(y,w) =0\  ,\ \ \ \  i=1,2\  .
\end{equation}
Let the Wronskian be
\begin{equation}\label{Eq12}
W(w)=W(\tilde \psi_1, \tilde \psi_2 )= \tilde \psi_1 \tilde \psi_{2,y} - 
\tilde \psi_2 \tilde \psi_{1,y}\  ,
\end{equation}
where $W(w)$ is $y$-independent.  
Using 
the two solutions
$\tilde \psi_1$ and $\tilde \psi_2$, the black- hole Green's function
can be expressed as
\begin{equation}\label{Eq13}
\tilde G(y,x;w) =- {1 \over {W(w)}} 
\left\{ \begin{array}{l@{\quad,\quad}l}
\tilde \psi_1(y,w) \tilde \psi_2(x,w) & y<x \  , \\
\tilde \psi_1(x,w) \tilde \psi_2(y,w) & y>x \  .
\end{array} \right.
\end{equation}

In order to calculate $G(y,x;t)$ using Eq. (\ref{Eq10}), one may close the 
contour of integration into the lower half of the complex frequency plane.
Then, one finds three distinct contributions to $G(y,x;t)$ \cite{Leaver} :

\begin{enumerate}
\item {\it Prompt contribution}. This arises from the integral along the large
semi-circle. It is this part, denoted $G^F$, which propagates the {\it
high}-frequency response.  For large frequencies the Green's function
approaches to the one of a {\it massless} scalar-field on a flat
spacetime background.  This term contributes to the {\it short}-time
response and can be shown to be effectively zero beyond a certain
time. Thus, it is not relevant for the late-time behaviour of the
field.

\item {\it Quasinormal modes}. These arise from the distinct singularities of
$\tilde G(y,x;w)$ in the lower half of the complex $w$-plane and is
denoted by $G^Q$. These singularities occur at frequencies for which
the Wronskian (\ref{Eq12}) vanishes. $G^Q$ is just the sum of the
residues at the poles of $\tilde G(y,x;w)$. Since each mode has
Im$w<0$ it decays {\it exponentially} with time.
\item {\it Tail contribution}. As will be shown 
later 
the intermediate asymptotic tail is
associated with the existence of a branch cut (in $\tilde \psi_2$)
placed along the interval $-m \leq w \leq m$. This tail arises from
the integral of $\tilde G(y,x;w)$ around the branch cut (denoted by
$G^C$) 
which 
leads to an {\it oscillatory} inverse {\it power-law}
behaviour of the field. 
Since we are interested in  the intermediate asymptotic behaviour of a 
SI scalar-field  our goal is to evaluate $G^C(y,x;t)$.              
\end{enumerate}


\section{The intermediate asymptotic behaviour of a self-interacting scalar-field}\label{Sec4}

\subsection{The $M \ll r \ll {M / {(Mm)^{2}}}$ approximation}\label{Sec4A}

It is well known that the late-time behaviour of {\it massless} fields
is determined by the backscattering from asymptotically {\it far}
regions \cite{Thorne,Price,HodPir1}. Thus, the late-time tails of
massless fields are dominated by the {\it low}-frequencies
contribution to the Green's function, for only low frequencies will be
backscattered by the small spacetime curvature or by the small
electromagnetic interaction at these asymptotic regions.  On the other
hand, it is also well known that {\it massive}-tails exist even in a
{\it flat} spacetime \cite{Morse}. This phenomena is related to the
fact that different frequencies forming a massive wave packet have
different phase velocities.  As will be shown in this paper, the
intermediate asymptotic behaviour of a {\it SI} scalar-field on a
Reissner-Nordstr\"om background is dominated by {\it flat} spacetime
effects. Namely, at intermediate times 
the backscattering from asymptotically {\it far}
regions (which dominates the tails of massless fields) 
is 
{\it negligible} compared to the flat spacetime massive tails that appear  here. 

Let us assume that both the observer and the initial data are situated
far away from the black-hole. We expend the wave-equation (\ref{Eq11})
for the SI scalar-field (in the field of the black-hole) as a power
series in $M/r$ and $Q/r$ and obtain (neglecting terms of order
$O[({{Mm} \over r})^{2}]$ and
higher) 
\begin{equation}\label{Eq14}
  \left[ {{d^2} \over {dr^2}} +w^{2}-m^{2} +{{4Mw^{2}-2Mm^{2}} \over r} - 
{{l(l+1)} \over {r^{2}}} \right ] \xi =0\  ,
\end{equation}
where $\xi=\lambda \tilde \psi$.  The term proportional to $M/r$
represent the Newtonian potential.  If we further assume that both the
observer and the initial data are situated in the region $r \ll {M
/ {(Mm)^{2}}}$ (and $M \ll r$), and we are interested in the
intermediate asymptotic behaviour of the field ($r \ll t \ll {M
/ {(Mm)^{2}}}$)
we can further approximate
Eq. (\ref{Eq14}) by
\begin{equation}\label{Eq15}
  \left[ {{d^2} \over {dr^2}} +w^{2}-m^{2} - 
{{l(l+1)} \over {r^{2}}} \right ] \xi =0\  ,
\end{equation}
Replacing Eq. (\ref{Eq14}) with Eq. (\ref{Eq15}) 
means that 
we 
{\it neglect} the backscattering of the field from
asymptotically {\it far} regions. 
Thus, the intermediate asymptotic
behaviour of {\it SI} scalar perturbations on a black-hole (or a
star) background depends only on the
field's parameters (namely, on the mass of the field) 
and 
it does 
{\it not} 
depend 
on the spacetime parameters.
The validity of this conclusion is verified by numerical simulations (see
Sec. \ref{Sec5}).

Let us now introduce a second auxiliary field $\tilde \phi$ defined by
\begin{equation}\label{Eq16}
\xi =r^{l+1} e^{i\sqrt{w^{2}-m^{2}} r} \tilde \phi (z)\  ,
\end{equation}
where
\begin{equation}\label{Eq17}
z=-2iwr\  .
\end{equation}
$\tilde \phi$(z) satisfies the confluent hypergeometric equation
\begin{equation}\label{Eq18}
\left[ z {{d^{2}} \over {dz^{2}}} +(2l+2-z) {d \over {dz}} -(l+1)
\right] \tilde \phi(z)=0\  .
\end{equation}
The 
two basic solutions required in order to build the black-hole
Green's function are (for $|w| \leq m$)
\begin{equation}\label{Eq19}
\tilde \psi_1 =Ar^{l+1} e^{-\varpi r} M(l+1,2l+2,2\varpi r)\  , 
\end{equation}
and
\begin{equation}\label{Eq20}
\tilde \psi_2 =Br^{l+1} e^{-\varpi r} U(l+1,2l+2,2\varpi r)\  ,
\end{equation}
where $\varpi = \sqrt{m^2 - w^2}$. $A$ and $B$ are normalization
constants. 
$M(a,b,z)$ and $U(a,b,z)$ are 
the two standard solutions to the confluent hypergeometric equation 
\cite{Abram}. $U(a,b,z)$ is a many-valued function, i.e. there 
is 
a cut in $\tilde \psi_2$.  Using Eqs. 13.6.6 and 13.6.24 of
\cite{Abram} one may write these solutions in a more familiar form
\begin{equation}\label{Eq21}
\tilde \psi_1 ={1 \over 2}Ar^{{1 \over 2}} \Gamma(l+ {3 \over 2}) ({1 \over 2}
\varpi )^{-(l+{1 \over 2})} I_{l+{1 \over 2}} (\varpi r)\  , 
\end{equation}
and
\begin{equation}\label{Eq22}
\tilde \psi_2 =\pi^{-{1 \over 2}} Br^{{1 \over 2}} (2 \varpi)^{-(l+{1
\over 2})} K_{l+{1 \over 2}} (\varpi r)\  ,
\end{equation}
where $I_{l+{1 \over 2}}$ and $K_{l+{1 \over 2}}$ are the modified
Bessel functions.
  
Using Eq. (\ref{Eq10}), one finds that the branch cut contribution to the 
Green's function is given by
\begin{equation}\label{Eq23}
G^C(y,x;t)={1 \over {2\pi}} \int_{-m}^{m} \tilde \psi_1(x,\varpi) \left[
{{\tilde \psi_2(y,\varpi e^{\pi i})} \over {W(\varpi e^{\pi i})}} -
{{\tilde \psi_2(y,\varpi )} \over {W(\varpi )}} \right] e^{-iwt} dw\  .
\end{equation}
For simplicity we assume that the initial data has a considerable support 
only for $r$-values which are smaller than the observer's location.
This, of course, does not change the {\it late}-time
behaviour. 

Using Eqs. 9.6.30 and 9.6.31 of \cite{Abram}, one finds
\begin{equation}\label{Eq24}
\tilde \psi_1(r,\varpi e^{\pi i})= \tilde \psi_1(r,\varpi )\  ,
\end{equation}
and
\begin{equation}\label{Eq25}
\tilde \psi_2(r,\varpi e^{\pi i})=- \tilde \psi_2(r,\varpi)+
{B \over A}{{ \pi^{1 \over 2} (-1)^{l+1}2^{-2l}} \over 
{\Gamma(l+{3 \over 2})}} 
\tilde \psi_1(r,\varpi)\  .
\end{equation}
Using Eqs. (\ref{Eq24}) and (\ref{Eq25}) it is easy to see that
\begin{equation}\label{Eq26}
W(\varpi e^{\pi i})=-W(\varpi)\  .
\end{equation}
From which 
we obtain the relation
\begin{equation}\label{Eq27}
{{\tilde \psi_2(r,\varpi e^{\pi i})} \over {W(\varpi e^{\pi i})}} -
{{\tilde \psi_2(r,\varpi )} \over {W(\varpi )}} =
{B \over A} {{\pi^{1 \over 2} (-1)^{l}2^{-2l}} \over {\Gamma(l+{3
      \over 2})}}
{{\tilde \psi_1(r,\varpi)} \over {W(\varpi)}}\  .
\end{equation}
Since $W(\varpi)$ is $r$-independent, we may use the $\varpi r \to 0$
asymptotic expansions of the Bessel functions (given by Eqs. 9.6.7 and
9.6.9 in \cite{Abram}) in order to evaluate it. One finds
\begin{equation}\label{Eq28}
W(\varpi)={-{1 \over 4}} AB \pi^{-{1 \over 2}} (2l+1) 
\Gamma (l+{1\over 2}) \varpi ^{-(2l+1)}\  .
\end{equation}
Finally, 
substituting (\ref{Eq27}) and (\ref{Eq28}) in (\ref{Eq23}) we obtain
\begin{equation}\label{Eq29}
G^C(y,x;t)={{(-1)^{l+1} 2^{2-2l}} \over {A^{2} (2l+1) \Gamma(l+{1
      \over 2}) \Gamma(l+{3 \over 2})}} 
\int_{0}^{m} \tilde \psi_1(y,\varpi) 
\tilde \psi_1(x,\varpi) \varpi ^{2l+1} e^{-iwt} dw\  .
\end{equation}

\subsection{Intermediate asymptotic at a fixed radius}\label{Sec4B}

It is easy to verify that 
in the large $t$ limit 
the effective contribution to the integral
in (\ref{Eq29}) 
arise 
from $|w|$=$O(m-{1 \over t})$ 
or equivalently $\varpi = O(\sqrt {{m \over t}})$. 
This is due to the rapidly oscillating term $e^{-iwt}$ which
leads to a mutual cancelation between the positive and the negative
parts of the integrand.  
In 
order to obtain the intermediate asymptotic
behaviour of the field at a fixed radius (where $x,y
\ll t$), we may use the $\varpi r \ll 1$ limit of $\tilde
\psi_1(r,\varpi )$. Using Eq. 9.6.7 from \cite{Abram} one finds
\begin{equation}\label{Eq30}
\tilde \psi_1(r,\varpi) \simeq {1 \over 2}Ar^{l+1}\  .
\end{equation}
We 
obtain
\begin{equation}\label{Eq31}
G^C(y,x;t)=
{{(-1)^{l+1} \pi^{{1 \over 2}} m^{l+1}} \over {2^{l} (2l+1) 
\Gamma(l+{1\over 2})}} (xy)^{l+1} t^{-(l+1)} J_{l+1}(mt)\  ,
\end{equation}
which in the $t \gg m^{-1}$ limit becomes
\begin{equation}\label{Eq32}
G^C(y,x;t)=
\sqrt{{2 \over \pi}}
{{(-1)^{l+1}} \over {(2l+1)!!}} m^{l+{1 \over 2}} (xy)^{l+1}
t^{-(l+{3 \over 2})} cos[mt-({1 \over 2}l+ {3 \over 4}) \pi ]\  .
\end{equation}

Thus, the intermediate asymptotic behaviour of the SI field at a fixed
radius is dominated by an {\it oscillatory} inverse {\it power-law} tail.

\subsection{Intermediate asymptotic along the black-hole outer horizon}
\label{Sec4C}

Next, we consider the behaviour of the SI scalar-field at the
black-hole outer-horizon $r_+$. While Eqs. (\ref{Eq21}) and
(\ref{Eq22}) are 
approximate 
solutions to the wave-equation
(\ref{Eq11}) in the $M \ll r \ll {M / {(Mm)^{2}}}$ region, they do
not represent the solution near the horizon. As $y \to -\infty$ the
wave-equation (\ref{Eq11}) can be approximated by the equation
\begin{equation}\label{Eq33}
\tilde \psi _{,yy} + w^2 \tilde \psi =0\  .
\end{equation}
Thus, we 
chose 
\begin{equation}\label{Eq34}
\tilde \psi_1(y,w)=C(w)e^{-iwy}\  ,
\end{equation}
and we use the form (\ref{Eq30}) for $\tilde \psi_1(x,w)$. In order to
match the $y \ll -M$ solution 
to 
the $y \gg M$ solution we assume
that the two solutions have the same temporal dependence. 
This assumption has been proven to be very successful for massless        
neutral \cite{Gundlach} and charged \cite{HodPir1,HodPir2} perturbations. 
In other words we 
assume that 
$C(w)$ 
is 
$w$-independent.  In this case one
should replace the roles of $x$ and $y$ in Eqs. (\ref{Eq23}) and
(\ref{Eq29}).  Using (\ref{Eq29}), we obtain
\begin{equation}\label{Eq35}
G^C(y,x;t)=
\Gamma_{0} \sqrt{{2 \over \pi}}
{{(-1)^{l+1}} \over {(2l+1)!!}} m^{l+{1 \over 2}} y^{l+1}
v^{-(l+{3 \over 2})} cos[mv-({1 \over 2}l+ {3 \over 4}) \pi ]\  ,
\end{equation}
where $\Gamma_{0}$ is a constant.

Thus, the intermediate asymptotic behaviour of the SI 
field along the black-hole 
outer-horizon is dominated by
an {\it oscillatory} inverse {\it power-law} tail. 

\section{Numerical Results}\label{Sec5}

\begin{figure}
\centering
\noindent
\includegraphics[width=15cm]{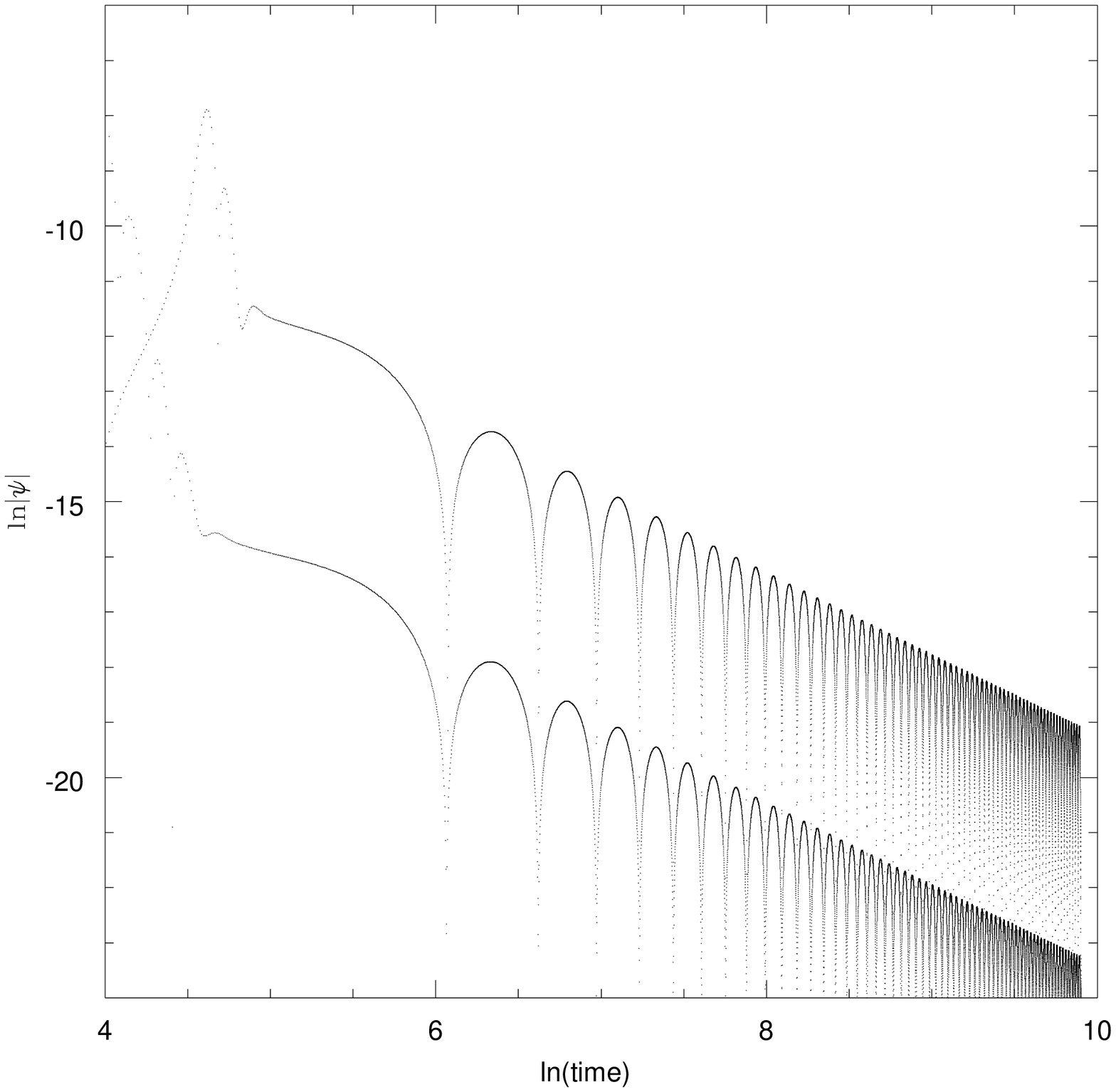}

\caption[l=0 mode]{\label{Fig1}
Evolution of the SI field $|\psi|$ on a Reissner-Nordstr\"om
background, for $l=0, M=0.5, Q=0.45$ and $m=0.01$. The initial data is
a Gaussian distribution with $v_{0}=50$ and $\sigma =2$.  The field at
a fixed radius ($y=50$) is shown as a function of $t$.
Along the black-hole outer horizon the field is shown as a function of
$v$.  The oscillatory {\it power-law} fall off is manifest at
intermediate times.  The power-law exponents are $-1.48$ at a fixed 
radius (top curve) and $-1.47$ along the black-hole outer-horizon 
(bottom curve).
These values are to be compared with the {\it analytically} predicted
value of $-1.5$.  The period of the oscillations is $T= \pi /m$ to
within 0.1\%, in agreement with the predicted value.}
\end{figure}

\begin{figure}
\centering
\noindent
\includegraphics[width=15cm]{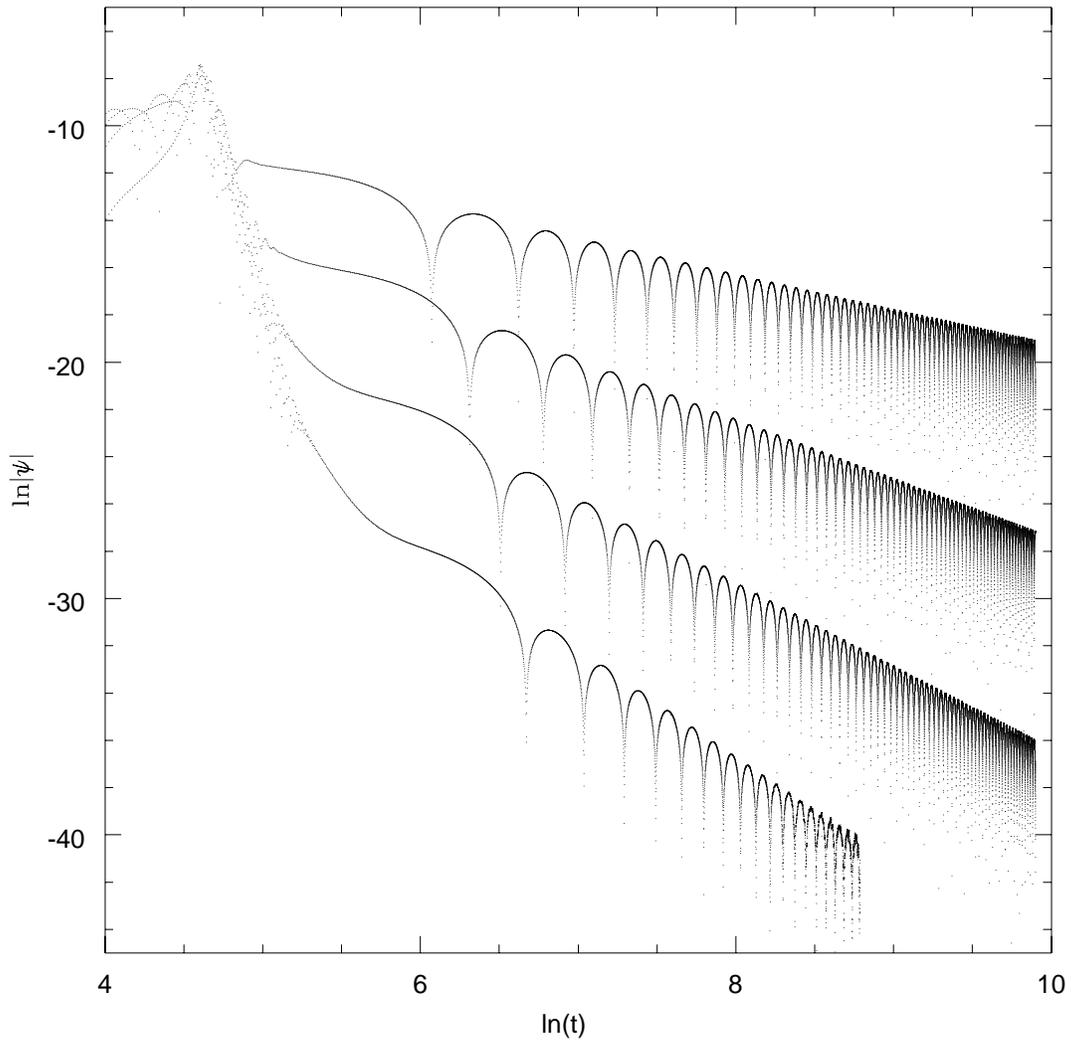}
\caption[various values of $l$]{\label{Fig2}
The amplitude of the field $|\psi (y=50,t)|$ for different multipoles
$l$=0,1,2 and 3 (from top to bottom).  The power-law exponents are
$-1.49, -2.50, -3.50$ and $-4.51$, respectively, in excellent
agreement with the {\it analytically} predicted values of $-1.5, -2.5,
-3.5$ and $-4.5$.  The oscillations period is $T= \pi /m$ to within
0.1\%, in agreement with the predicted value. The initial data are
those of Fig. \ref{Fig1}.}  \end{figure}

It is straightforward to integrate Eq. (\ref{Eq4}) using the method
described in \cite{Gundlach}. The late-time evolution of a SI
scalar-field is independent of the form of the initial data used. The
results presented here are 
for 
a Gaussian pulse on $u=0$
\begin{equation}\label{Eq36}
\psi (u=0,v)= A exp \left \{-\left [ (v-v_{0})/ \sigma \right ]^{2}
\right \}\  ,
\end{equation}
where the amplitude $A$ is physically irrelevant due to the linearity
of Eq. (\ref{Eq4}).
It should be noted that the evolution equation (\ref{Eq4}) is
invariant under the rescaling
\begin{equation}\label{Eq37}
r \to ar \  ,\  t \to at \  ,\  M \to aM \  ,\  Q \to aQ\  ,\  m \to m/a \  ,
\end{equation}
where $a$ is some positive constant.  The black-hole mass and charge
are set equal to $M=0.5$ and $Q=0.45$, respectively. We have chosen 
an initial field-profile with 
$m=0.01\ ,v_{0}=50$ and $\sigma =2$. We studied the behaviour of the
field $\psi$ at a fixed radius
and along the black-hole outer horizon (approximated by the null surface
$u=u_{max}$, where $u_{max}$ is the largest value of $u$ on the grid).
The numerical results for the $l=0$ mode are shown in Fig. \ref{Fig1}.
Initially, the evolution is dominated by the prompt contribution and
the quasinormal ringing. However, at intermediate times a definite oscillatory
{\it power-law} fall off is manifest.  The power-law exponents are
$-1.48$ at a fixed radius (top curve) and $-1.47$ along the black-hole
outer-horizon (bottom curve).  These values are to be compared with
the {\it analytically} predicted value of $-1.5$, see
Eqs. (\ref{Eq32}) and (\ref{Eq35}). The period of the oscillations
(for $|\psi|$) 
equals 
$T= \pi /m$ to within 0.1\%, again in
agreement with the predicted value.  
Fig. \ref{Fig2}. depicts 
the dependence of the intermediate asymptotic tails (at a fixed
radius $y=50$) on the multipole
index $l$.  The numerical values of the power-law exponents,
describing the fall-off of the field at intermediate times (after a
period of quasinormal ringing) are $-1.49, -2.50, -3.50$ and $-4.51$
for $l$=0,1,2 and 3, respectively.  These numerical values are in
excellent agreement with the {\it analytically} predicted values of
$-1.5, -2.5, -3.5$ and $-4.5$.  Again, the period of the oscillations
is 
$T= \pi /m$ to within 0.1\%, in agreement with the
predicted value.

\begin{figure}
\centering
\noindent
\includegraphics[width=15cm]{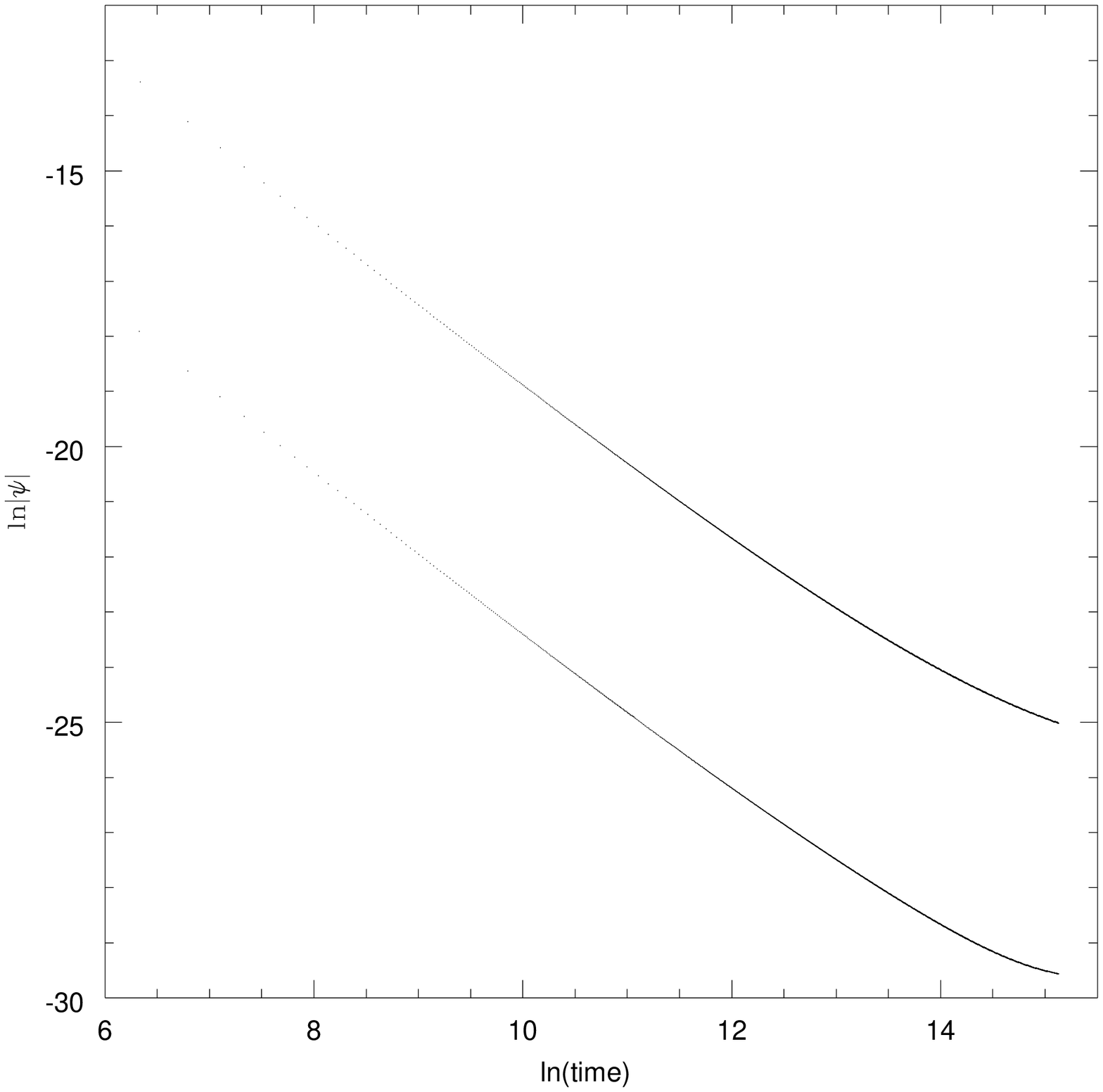}
\caption[complicated evolution]{\label{Fig3}
Late-time evolution of the SI field $|\psi|$ on a Reissner-Nordstr\"om
background. The field at future timelike infinity
($y=50$) is shown as a function of $t$ (top curve). Along the
black-hole outer horizon the field is shown as a function of $v$
(bottom curve). 
For the clarity of the figure 
we display only the maximas of the oscillations (the field 
oscillates with a period of $T= \pi /m$ to within 0.2\%).
At {\it late-}times the field's amplitude 
decays {\it slower} than any
power-law. The initial data are those of Fig. \ref{Fig1}.}
\end{figure}

The analytical derivations and their numerical confirmations presented
so far are restricted to the {\it intermediate} asymptotic regime. 
The {\it late-}time evolution of the field (for the $l=0$ mode)
is shown in Fig. \ref{Fig3} (for the clarity of the figure 
we display only the maximas of the oscillations). 
The initial data are those of Fig. \ref{Fig1}.
Shown are the behaviour of the field
$|\psi|$ along the asymptotic regions of timelike infinity $i_{+}$
(top curve) and along the black-hole outer horizon (bottom curve). It
is clear that the field's amplitude {\it decays} with time, in
agreement with the no-hair theorem. The decay rate is {\it slower}
than any power-law.  Again, we find that the field $|\psi|$ oscillates
with a period of $T= \pi /m$ to within 0.2\%.
We have found that the larger is the field's mass the sooner it leaves
the intermediate asymptotic phase of an inverse power-law decay.

\section{Summary and physical implications}\label{Sec6}
We have studied analytically the intermediate 
asymptotic evolution of a {\it SI} 
scalar-field on a Reissner-Nordstr\"om background. 
Following the {\it no-hair theorem} we have focused attention
on the physical mechanism by which a {\it SI} hair decays. 
The main results and their physical implications are:


{\it Oscillatory} inverse {\it power-law} tails develop at
intermediate times at a fixed radius 
and along the black-hole outer horizon (as long as the
initial data has a considerable support only in the region $M \ll r
\ll {M \over {(Mm)^{2}}}$. Actually, our analytical derivations hold
even in the case $M \ll r \ll {{Ml(l+1)} \over {(Mm)^{2}}}$ for
$l>0$).
The dumping exponents, describing the fall-off of a {\it SI} field at
intermediate times, are {\it smaller} compared with those of {\it massless}
neutral perturbations \cite{Price,Gundlach}. For $l > {2 \over
\sqrt{3}}|eQ| - {1 \over 2}$ (where $e$ is the field's charge) these
dumping exponents are also smaller than those of {\it massless}
charged perturbations \cite{HodPir2}. 
While the asymptotic behaviour of {\it massless}
perturbations is dominated by backscattering from asymptotically far
regions [and thus depends on the {\it spacetime} parameters, $M$ (for
neutral perturbations) and $Q$ (for charged perturbations)], the 
intermediate asymptotic behaviour of a SI field depends on the {\it field's}
parameters (namely, on the field's mass $m$). In other words, dealing
with SI perturbations, one may {\it neglect} the backscattering from
asymptotically far regions at intermediate times.

Using the numerical scheme we have shown that 
at {\it late-}times the inverse power-law decay is
replaced by another pattern of decay, which is {\it slower} than any power-law.
This late-time behaviour deserves a further analytic study.
The late-time behaviour of SI perturbations implies that a black-hole
which forms from a gravitational collapse of SI fields becomes
``bald'' {\it slower} than one which forms during a gravitational
collapse of massless fields.
Since {\it SI} perturbation fields decay at late-times slower than any
power-law they are expected to cause a {\it mass-inflation}
singularity during a gravitational collapse which leads to the
formation of a black-hole.  Moreover, SI fields are expected to {\it
dominate} the mass-inflation phenomena during a gravitational collapse
(this is caused by the {\it slower} decay of SI perturbations compared
with massless ones).



We have confirmed our {\it analytical} results by numerical
calculations, showing that the intermediate asymptotic behaviour of a
SI field is dominated by an {\it oscillatory inverse power-law} tail 
(with the analytically predicted dumping exponents and
oscillation frequency).


\bigskip
\noindent
{\bf ACKNOWLEDGMENTS}
\bigskip

This research was supported by a grant from the Israel Science Foundation.

\begin{thebibliography}{99}
\bibitem{Price}R.H. Price, Phys. Rev. {\bf D5}, 2419 (1972).

\bibitem{HodPir1} S. Hod and T. Piran, gr-qc/9712041, to be published 
in Phys. Rev. 
{\bf D}.

\bibitem{HodPir2} S. Hod and T. Piran, gr-qc/9801001, to be published 
in Phys. Rev.
{\bf D}.

\bibitem{Poisson}E. Poisson and W. Israel, Phys. Rev. {\bf D41}, 1796 (1990).

\bibitem{Burko}L. M. Burko, in Abstracts of contributed papers 
to the 15th conference
on General Relativity and Gravitation, Pune, 1997 (unpublished).

\bibitem{Leaver} E.W. Leaver, Phys. Rev. {\bf D34}, 384 (1986).

\bibitem{Thorne} K.S. Thorne, p. 231 in Magic without magic: John Archibald
Wheeler Ed: J.Klauder (W.H. Freeman, San Francisco 1972).

\bibitem{Morse}P.M. Morse and H. Feshbach, Methods of
theoretical physics (McGraw-Hill, New York 1953).

\bibitem{Abram} M. Abramowitz and I.A. Stegun, Handbook of mathematical
functions (Dover Publications, New York 1970).

\bibitem{Gundlach}C. Gundlach, R.H. Price, and J. Pullin, Phys. Rev. {\bf D49},
883 (1994).

\end{thebibliography}
\end{document}